\documentclass[11pt,a4paper]{article}
\usepackage{jcappub}
\usepackage{graphicx}
\usepackage{dcolumn}
\usepackage{bm}
\usepackage{color}

\input{colordvi.tex}

\newcommand{\beq}{\begin{equation}}
\newcommand{\eeq}{\end{equation}}
\newcommand{\beqa}{\begin{eqnarray}}
\newcommand{\eeqa}{\end{eqnarray}}

\newcommand{\GeV}{~\mbox{\rm GeV}}

\newcommand{\dd}{\mathrm{d}}
\newcommand{\Mpl}{M_{\rm Pl}}
\newcommand{\Mpc}{{\rm Mpc}}
\newcommand{\G}{{\rm G}}
\newcommand{\pc}{{\rm pc}}
\newcommand{\mcA}{\mathcal{A}}

\begin{document}

\title{
Large-scale magnetic fields can explain the baryon asymmetry of the Universe
}
\author[a]{Tomohiro Fujita}
\author[b]{, Kohei Kamada}

\affiliation[a]{Stanford Institute for Theoretical Physics and Department of Physics,\\
Stanford University, Stanford, CA 94306, USA}

\affiliation[b]{Department of Physics and School of Earth and Space Exploration,
\\ Arizona State University, Tempe, AZ 85287, USA}

\emailAdd{tomofuji@stanford.edu}
\emailAdd{kohei.kamada@asu.edu}


\abstract{
Helical hypermagnetic fields in the primordial Universe can produce the observed amount of baryon asymmetry through the chiral anomaly without any ingredients beyond the standard model of particle physics.
While they generate no $B-L$ asymmetry, the generated baryon asymmetry survives the spharelon washout effect, because the generating process remains active until the electroweak phase transition. 
Solving the Boltzmann equation numerically and finding an attractor solution, we show that the baryon asymmetry of our Universe can be explained, if the present large-scale magnetic fields indicated by the blazar observations have a negative helicity and existed in the early Universe before the electroweak phase transition. We also derive the upper bound on the strength of the helical magnetic field, which is tighter than the cosmic microwave background constraint, to avoid the overproduction of baryon  asymmetry.
}
\maketitle

%
%
%

\section{Introduction}

The origin of the baryon asymmetry of the present Universe is one of the biggest problems in both cosmology and high energy physics. 
The prevailing lore is that the baryon asymmetry is almost impossible to be generated within the standard model of particle physics (SM) because 
it is hard to satisfy Sakharov's conditions \cite{Sakharov:1967dj}. 
Therefore, it is often explored by assuming some extensions of the SM. 

However, a remarkable mechanism is studied in Ref.~\cite{Giovannini:1997eg} and  recently revisited  in Ref.~\cite{Anber:2015yca} in the context of the pseudoscalar inflation model.
In this mechanism, baryon asymmetry is generated from 
 helical magnetic fields thorough the chiral anomaly in the SM $U(1)_Y$ gauge interaction. 
(See also other studies of baryogenesis and magnetic fields~\cite{Joyce:1997uy, Brustein:1998du,Giovannini:1999wv,Giovannini:1999by,Bamba:2006km,Bamba:2007hf,Boyarsky:2011uy,Dvornikov:2011ey,Boyarsky:2012ex,Semikoz:2012ka,Tashiro:2012mf,Dvornikov:2012rk,Long:2013tha,Dvornikov:2013bca,Semikoz:2013xkc,Sabancilar:2013raa,Bhatt:2015ewa,Semikoz:2015wsa,Boyarsky:2015faa,Zadeh:2015oqf,Long:2016uez}.)
From the chiral anomaly, baryon asymmetry can be generated if there exists a time-varying helicity of the hypermagnetic fields. 
Although the hypermagnetic helicity is a good conserved quantity in the early Universe, it slightly changes with time due to the large but finite conductivity of the Universe. Here, the time-varying helical hypermagnetic fields breaks spontaneously $T$ symmetry as well as $C$ and $CP$ symmetry, and  the baryon asymmetry is generated without the strong departure from  thermal equilibrium like spontaneous baryogenesis \cite{Cohen:1987vi}. 
It should be noted that the mechanism itself does not require any ingredients beyond the SM, which motivates us to explore this mechanism further.%
\footnote{Exactly speaking, Ref.~\cite{Anber:2015yca} considers   a physics beyond the SM,  namely, a pseudoscalar inflation model with a dimension-five coupling term between the pseudoscalar and the $U(1)_Y$ gauge field, to generate the helical magnetic field and derives a constraint on the model due to the overproduction of baryon asymmetry. However, no effect beyond the SM is involved in the generating process of the baryon asymmetry.}

We study the mechanism from an opposite side to Ref.~\cite{Anber:2015yca}
in the following way; whereas Ref.~\cite{Anber:2015yca} studies the generation of baryon asymmetry in a forward-in-time way from a specific magnetogenesis mechanism, namely, pseudoscalar inflation, we study it in a backward-in-time way from the present cosmic magnetic fields, without specifying the magnetogenesis mechanism. 
In this sense, our study is complementary to the study in Ref.~\cite{Anber:2015yca}. 

Furthermore, compared to Ref.~\cite{Anber:2015yca}, we additionally take into account the following points; (i) the constraints on the cosmic magnetic fields imposed by observations; (ii) the nontrivial evolution of the magnetic field governed by the magnetohydrodynamic effect, and (iii) the contribution from the Yukawa interaction in the Boltzmann equation.

The upper and lower bounds on the strength of the present large-scale magnetic fields $B_0$ are given as $10^{-17} \G \lesssim B_0 \lesssim 10^{-9} \G$ by the observations of the cosmic microwave background (CMB)~\cite{Ade:2015cva} and the gamma rays from blazars~\cite{Neronov:1900zz,
Tavecchio:2010mk,Ando:2010rb,Dolag:2010ni,Essey:2010nd,Taylor:2011bn,
Takahashi:2013uoa,Chen:2014rsa}, respectively. 
In particular, it should be remarked that the latter indicates the existence of the 
large-scale magnetic fields.
Provided that the magnetic fields are helical and generated before the electroweak phase transition, they must produce some baryon asymmetry via the chiral anomaly. Indeed, it is claimed that the diffuse gamma ray observation infers a nonvanishing helicity of the present large-scale magnetic field~\cite{Tashiro:2013ita, Tashiro:2014gfa, Chen:2014qva}.
Reconstructing the properties of the magnetic fields in the 
early Universe from these observational results,
we entirely explore the amount of produced baryon asymmetry in the allowed parameter region of the present magnetic fields.

In order to estimate the resultant baryon asymmetry qualitatively, we also take into account 
the evolution of the magnetic field governed by the magnetohydrodynamic effect.
It is known that the time evolution of magnetic fields in the Universe is nontrivial in general, and it is not necessarily the adiabatic evolution in which the physical strength of the magnetic field decays in proportional to $a^{-2}(t)$, where $a(t)$ is the scale factor. 
This is because the magnetohydrodynamical effects may 
cause the inverse cascade process, which we will describe in Sec.~\ref{sec2}.  
(For recent review on magnetohydrodynamics (MHD) in astrophysics, 
see {\it e.g.} Ref.~\cite{Brandenburg:2004jv, Durrer:2013pga}). 
Fortunately, the time evolution of some properties of the magnetic field, namely the peak strength and the correlation length, can be estimated by an analytical method~\cite{Muller, Banerjee:2004df,Durrer:2013pga}, which enables us to evaluate the properties of magnetic fields around the electroweak scale from the present observations.
Based on this analytical estimate, we will see that the observed baryon asymmetry can be explained if the magnetic field  
have been undergoing the inverse cascade process above the electroweak scale. 
Note that the adiabatic evolution is applicable to relatively weaker magnetic fields with longer correlation length.  However, It will be shown that only negligible baryon asymmetry can be produced in that case for the observationally allowed present strength of the magnetic field. 
Therefore it is crucial to take into account the MHD effect in the study of baryogenesis from helical magnetic fields in our approach.

We here comment on the effect of the Yukawa interaction. 
It is often discussed that for the present baryon asymmetric Universe, $B-L$ asymmetry 
($B$ and $L$ are baryon and lepton numbers, respectively.) 
must be generated, otherwise the $B+L$ violating sphaleron process washes out the 
baryon asymmetry even if $B+L$ asymmetry is produced. 
However, this washout process is effective when the sphaleron process 
as well as all the Yukawa interactions are in equilibrium~\cite{Harvey:1990qw}. Therefore, in order to evaluate the resultant baryon asymmetry qualitatively, 
not only the sphaleron 
process but also the Yukawa interaction should be taken into account.
We solve the Boltzmann equations which include the Yukawa interactions as well as
the chiral anomaly and the spharelon effects simultaneously. While the equations are fairly complicated and require numerical calculations, we find an attractor behavior of the generated baryon asymmetry. The attractor appears when the source of the baryon asymmetry from the helical hypermagnetic field and the washout effect through the weakest (electron) Yukawa interaction are balanced. We derive a simple analytical expression of the resultant baryon asymmetry which shows an excellent agreement with the numerical result. The attractor is so strong that the baryon asymmetry depends only on the size of the source term at the electroweak phase transition and the other parameters ({\it e.g.} the initial temperature at which the source becomes effective) are irrelevant.

In this paper,  we find that the baryon asymmetry in our Universe can be explained  by magnetic fields with present strength  $10^{-14} \G \lesssim B_0 \lesssim 10^{-12} \G$ allowing for theoretical uncertainties,
if the magnetic fields have entered the 
inverse cascade regime before the electroweak phase transition. 
On the other hand, for $10^{-12} \G \lesssim B_0 \lesssim 10^{-9} \G$ the baryon asymmetry is basically overproduced and hence such strength of the present magnetic fields are disfavored. This problem can be avoided and the present baryon asymmetry can be explained if the magnetic fields evolve adiabatically before the electroweak
phase transition and enter the inverse cascade regime at a certain time after that. 
It should be noted that we do not specify the magnetogenesis mechanism in this paper and keep the analysis as general as possible. 
Our results give a further motivation of the study on magnetogenesis mechanisms that produce helical magnetic fields.\footnote{To the best of our knowledge, no mechanism is known to be able to produce magnetic fields which satisfy the observational lower bound, still less our scenario \cite{Vachaspati:2001nb,Copi:2008he,Demozzi:2009fu, Barnaby:2012tk,Fujita:2012rb,Fujita:2013qxa,Fujita:2014sna,Ferreira:2014hma}.
Therefore it is challenging and intriguing open question how the magnetic field are generated, while we do not explore it in this paper.}

This paper is organized as follows. 
In the next section, we discuss the evolution of magnetic fields, taking into account the inverse cascade process. We also describe the helicity conservation
and the constraints on the large-scale magnetic fields.
In Sec.~\ref{sec3},  we study the chiral anomaly in the SM and derive the evolution 
equation for the baryon asymmetry. An analytical expression for the attractor behavior is also given there. 
The quantitative results for the parameter space that can be responsible for 
the present baryon asymmetry obtained by numerical calculations are shown
in Sec.~\ref{sec4}. 
The final section is devoted to summary and discussion.

\section{Evolution of helical magnetic fields}
\label{sec2}

In this section we study the evolution of helical magnetic fields from the early Universe until the present and evaluate their properties. They will act as the source for the baryon asymmetry which we will discuss in more detail in the next section. Although we do not specify the generation mechanism of the helical magnetic fields, their time evolution can be generically obtained from their present properties such as the strength $B_0$ and the correlation length $\lambda_0$. 
Note that the electroweak gauge symmetry is restored at temperatures above the electroweak scale $T\simeq 10^2$ GeV. 
We here assume that hypermagnetic fields [$U(1)_Y$ gauge field] are
generated at an earlier time and 
transform into magnetic fields [$U(1)_{\rm EM}$ gauge field] at the electroweak phase transition.\footnote{
The possibilities of magnetogenesis at the electroweak phase transition are also discussed (see {\it e.g.} Ref.~\cite{Vachaspati:1991nm,Sigl:1996dm}). But we here assume that the electroweak phase transition does not significantly affect the evolution of the (hyper)magnetic fields except for the effect discussed above. }
Although a part of hypermagnetic fields transforms into $Z$ boson, the strength of  (hyper)magnetic field changes only around 10 \%~\cite{Dimopoulos:2001wx}. Thus hereafter we neglect the effect and the hypermagnetic field is called the magnetic field unless explicitly stated.

In our Universe, the magnetic fields and the plasma fluid of charged particles can be significantly coupled and their nonlinear interaction may govern their evolution. In that case, the magnetohydrodynamic (MHD) effect should be taken into account and the physical strength $B_p(t)$ and the physical correlation length $\lambda_B(t)$ of the magnetic field do not necessarily evolve adiabatically, $B_{p} \not\propto a^{-2}(t)$ or $\lambda_B \not\propto a(t)$, where $a(t)$ is the scale factor. 
For instance, the turbulence of the plasma fluid may cause the magnetic correlation scale $\lambda_B(t)$ to grow faster than $a(t)$. Therefore it is not trivial to obtain the precise evolution of the magnetic fields. In general, dedicated numerical simulations are needed to solve the nonlinear MHD equations. However, as we shall see in this section, the peak strength and the correlation length of the magnetic field can be estimated by an analytical method
which has been developed in the literature
\cite{Muller, Banerjee:2004df,Durrer:2013pga}.
In particular, the helicity of magnetic fields is a useful quantity
and it substantially helps the analytic estimation.

\subsection{Three different evolution scenarios}

The cosmic magnetic fields are affected by several different effects, the interaction with turbulent fluid, the viscous diffusion, the free streaming of photon and neutrino, etc~\cite{Banerjee:2004df}. For our purpose, however, we can focus on the effect of the turbulent plasma fluid.
Here we assume that the initial spectrum of the magnetic fields has a peak at $\lambda_B(t_{\rm ini})$ and is blue-tilted on the larger scales, and then the magnetic helicity can be evaluated at the time-dependent peak scale $\lambda_B(t)$ during the course of their evolution.
The coupling between the magnetic field and the turbulent fluid becomes relevant, if the typical scale of the turbulence $\lambda_T \simeq v_T t$ reaches the scale of the magnetic field $\lambda_B$, where $v_T$ is the velocity of the fluid and $t$ is the cosmic time. On the other hand, if the turbulence scale is negligible compared to the magnetic scale, $\lambda_B\gg \lambda_T$, the adiabatic evolution of the magnetic fields takes place, $\lambda_B(t) \propto a(t)$.  
Since the turbulence scale grows faster than the adiabatically evolving 
magnetic scale, the former eventually catches up to the latter. 
After $\lambda_B$ and $\lambda_T$ become comparable, 
the magnetic correlation length is synchronized with   $\lambda_T$, because the smaller scale part of the magnetic power spectrum is lost due to the interaction with the turbulence. 
In the developed turbulence, $v_T$ is comparable to the Alfv\'en velocity, $v_A(t) \equiv
B_{p}/\sqrt{\rho_{\rm ch}+p_{\rm ch}}$, where $\rho_{\rm ch}$ and $p_{\rm ch}$ are the energy density and the pressure of the charged particles interacting with the magnetic field.
One finds~\cite{Banerjee:2004df} (see also \cite{Durrer:2013pga,Caprini:2014mja}),
\begin{align}
\lambda_B &\simeq v_A t\simeq \frac{B_p}{2 H}\sqrt{\frac{3}{4\rho_{\rm ch}}} =\frac{45 B_p \Mpl/T^4}{2\pi^2 \sqrt{g_{*}^{\rm tot}g_{*}^{\rm ch}}}
\notag\\
&=2.6\times 10^{-29}\Mpc\left(\frac{g_*^{\rm tot}(T)}{106.75}\right)^{-\frac{1}{2}}\left(\frac{g_*^{\rm ch}(T)}{82.75}\right)^{-\frac{1}{2}}\left(\frac{B_p}{10^{20}\G}\right)\left(\frac{T}{10^2\GeV}\right)^{-4} ,
\label{turbulent eq}
\end{align}
where $M_{\rm pl}=2.43 \times 10^{18}$ GeV is the reduced Planck mass, $T$ is the temperature, and $g_{*}^{\rm tot}$ and $g_{*}^{\rm ch}$ are the number of degree of freedom of all the particles in the thermal bath and the U(1) charged particles, respectively. As we shall see soon in Eq.~\eqref{l case I}, $\lambda_B$ grows faster than the adiabatic case in this regime and this process is called the {\it inverse cascade}.

Depending on the time when the inverse cascade starts, we have the following three
different evolution scenarios of the magnetic fields. (i) The solely inverse cascade case: The magnetic fields undergo the inverse cascade right after their generation. (ii) The transition case: First the magnetic fields adiabatically evolve, and subsequently the inverse cascade starts at a temperature $T=T_{\rm TS}$. (iii) The solely adiabatic case: The magnetic fields always evolve adiabatically and never experience the inverse cascade process.

As we will see, the helical magnetic field which has undergone the inverse cascade process can produce large baryon asymmetry,
while the solely adiabatic case produces very little baryon asymmetry.
Therefore, we mainly discuss the case (i) and (ii). 

\subsection{Helicity conservation}

Since the equations obtained in the previous subsection give only a relationship between $B_p(T)$ and $\lambda_B(t)$, we need another relation to determine each of them.%
\footnote{However, in the solely adiabatic case (iii), it is trivial that $B_p(t)=a^{-2}(t)B_0$ and $\lambda_B(t)= a(t) \lambda_0$.}
Then it is useful to introduce the helicity of the magnetic field,
\begin{equation}
{\cal H} \equiv \int_V d^3 x \, \bm{A}\cdot\bm{\mathcal{B}}= 
\int_V d^3 x \epsilon_{ijk} A_i \partial_j A_k,
\label{H def}
\end{equation}
with $\bm{A}$ being the vector potential and $\bm{\mathcal{B}}\equiv \bm{\nabla}\times\bm{A}$. 
It is well known that the helicity  represents the breaking of the parity (see Appendix \ref{ap1}) and it
is an approximate 
conserved quantity for sufficiently large electrical conductivity $\sigma$ [see Eq.~\eqref{h dot 1}].
The helicity density averaged over the 
cosmological scales is also conserved, and 
we can estimate it in terms of characteristic physical strength $B_p$ and physical length of the magnetic field $\lambda_B$ as
\begin{equation}
h \equiv \lim_{V \to \infty} \frac{\cal H}{V} \ \simeq \  a^3(t) \lambda_B(t)B_p^2(t)\simeq {\rm const}. 
\label{helicity conservation}
\end{equation}
From the helicity conservation and the relation determined by the inverse cascade
process, we are now ready to determine the properties of magnetic field 
at a given temperature $T$. From
Eqs.~\eqref{turbulent eq} and \eqref{helicity conservation}, as well as the entropy conservation $g_{*s} a^3 T^3$=const., we find that $B_p$ and $\lambda_B$ in the case (i) or for $T<T_{\rm TS}$ in the case (ii) are given by
\begin{align}
B_p^{\rm IC} (T) &\simeq 9.3 \times10^{19}\G 
\left(\frac{T}{10^2\GeV}\right)^{7/3}
\left(\frac{B_0}{10^{-14}\G}\right)^{2/3}
\left(\frac{\lambda_0}{1\pc}\right)^{1/3}\mathcal{G}_B(T),
\label{B case I}
\\
\lambda_B^{\rm IC} (T) &\simeq 2.4\times 10^{-29} \Mpc 
\left(\frac{T}{10^2\GeV}\right)^{-5/3}
\left(\frac{B_0}{10^{-14}\G}\right)^{2/3}
\left(\frac{\lambda_0}{1\pc}\right)^{1/3}\mathcal{G}_\lambda(T).
\label{l case I}
\end{align}
where the superscript ``IC" represents that the magnetic fields undergo the inverse cascade process, and $\mathcal{G}_B(T)\equiv (g_{*}^{\rm tot}(T)/106.75)^{1/6}(g_{*}^{\rm ch}(T)/82.75)^{1/6}(g_{*s}(T)/106.75)^{1/3}$ and $\mathcal{G}_\lambda(T)\equiv (g_{*}^{\rm tot}(T)/106.75)^{-1/3}(g_{*}^{\rm ch}(T)/82.75)^{-1/3}(g_{*s}(T)/106.75)^{1/3}$ denote the weak dependence on the number of the degree of freedom.
The temperature dependences of $B_p$ and $\lambda_B$ of this analytic estimate coincide with the numerical results in Ref.~\cite{Saveliev:2013uva}.
In the case (ii), 
the strength and correlation length of magnetic field 
at $T>T_{\rm TS}$ are given by
\begin{align}
B_p^{\rm AD}(T) &\simeq B_p^{\rm IC}(T_{\rm TS} ) \left(\frac{g_{*s}(T)}{g_{*s}(T_{\rm TS})}\right)^{2/3} \left(\frac{T}{T_{\rm TS}}\right)^2,
\label{B case II}
\\
\lambda_B^{\rm AD} (T) &\simeq \lambda_B^{\rm IC} (T_{\rm TS}) \left(\frac{g_{*s}(T_{\rm TS})}{g_{*s}(T)}\right)^{1/3} \left(\frac{T_{\rm TS}}{T}\right),
\label{l case II}
\end{align}
where the superscript ``AD" represents that the magnetic fields experience the transition from the adiabatic evolution into the inverse cascade regime.

It should be noted that the helicity (density) is not {\it completely} conserved. The large but finite electrical conductivity gives a slight time variation of helicity density, 
which will be important for baryogenesis. 
The time derivative of the helicity density is
\begin{equation}
\dot{h}=
\lim_{V\to \infty} \frac{2}{V} \int_V \dd^3 x \epsilon_{ijk} \dot{A}_i\partial_j A_k=- a^2 \frac{2}{\sigma} \left\langle\bm{B}\cdot \nabla \times \bm{B}\right\rangle
\simeq a^{3} \frac{4\pi }{\sigma} \frac{B_p^2}{\lambda_B},
\label{h dot 1}
\end{equation}
where we have used the Ampere's law and the generalized Ohm's law\footnote{Here we omit the chiral magnetic effect \cite{Vilenkin:1980fu,Vilenkin:1982pn}, since it gives 
only minor changes to the results and is negligible. }
and the bracket means that the quantity is averaged over the cosmological scales.%
\footnote{If the sign of the helicity is not uniform on the cosmological scale, the averaged value could be much smaller than Eq.~\eqref{h dot 1}.
For instance, however, the inflationary magnetogenesis model with the $\varphi F\tilde{F}$ coupling produces the helical magnetic fields with the uniform sign \cite{Anber:2006xt,Caprini:2014mja,Fujita:2015iga}.}
The sign is chosen for later convenience.\footnote{The negative helicity of the hypermagnetic fields leads the positive baryon asymmetry. If it is positive, negative baryon asymmetry will be generated.} We can see that ${\dot h}$ vanishes for $\sigma \rightarrow \infty$.  
At the same time, the validity of the helicity conservation can be confirmed if $|{\dot h}/hH| \ll 1$ is satisfied. 
We can evaluate it as
\begin{equation}
\left|\frac{\dot{h}}{H h}\right| \simeq   10^{-9} 
\left(\frac{T}{10^2 \GeV}\right)^{1/3}
\left(\frac{B_0}{10^{-14}\G}\right)^{-4/3}
\left(\frac{\lambda_0}{1\pc}\right)^{-2/3}\mathcal{G}_h(T),
\end{equation}
for the case (i), and smaller for the case (ii) and (iii). Here $\sigma\simeq 100T$ \cite{Baym:1997gq} and $\mathcal{G}_h(T)\equiv (g_{*}^{\rm tot}(T)/106.75)^{1/6}(g_{*}^{\rm ch}(T)/80)^{2/3}(g_{*s}(T)/106.75)^{-2/3}$ are used.  Thus we conclude that helicity has a nonzero time evolution but the helicity conservation is a very good approximation [see Eqs.~\eqref{presentMF} and \eqref{obs lower bound}]. 

\subsection{Constraints on the cosmic magnetic fields}

In this subsection, we shortly discuss several constraints on the magnetic fields in the Universe which should be appreciated in our scenario.
First, evaluating $\lambda_B\simeq v_A t$ at present, one finds the relation between the present strength $B_0$ and correlation length $\lambda_0$ 
of the magnetic field as~\cite{Banerjee:2004df,Durrer:2013pga,Caprini:2014mja}
\begin{equation}
\lambda_0 \simeq 10^{-6} \Mpc \left(\frac{B_0}{10^{-14} \G}\right). 
\label{presentMF}
\end{equation}
This relation should be applied to the magnetic fields which have experienced the inverse cascade process [i.e. the case (i) and (ii)].

Second, the observations of the gamma ray from blazars and the cosmic microwave background (CMB) give the lower and upper bound of the present strength of the magnetic field, respectively. The simultaneous GeV-TeV multi-wavelength observations of blazars infer the lower bound of the present strength of the magnetic field as~\cite{Taylor:2011bn}
(see also \cite{Neronov:1900zz,Tavecchio:2010mk,Ando:2010rb,Dolag:2010ni,Essey:2010nd,Takahashi:2013uoa,Chen:2014rsa})
\begin{equation}
B_0  \gtrsim  10^{-17} \G
\times\left\{
\begin{array}{cc}
\left(\lambda_0/1 {\rm Mpc} \right)^{-1/2} &  (\lambda_0<1{\rm Mpc})\\
 1 & (\lambda_0>1{\rm Mpc})
\end{array}\, \right.
.
\label{obs lower bound}
\end{equation}
 On the other hand, the observation of the CMB temperature anisotropy puts the upper bound on the current strength as $B_0\lesssim 10^{-9}\G$ on the CMB scales $\lambda_0\gtrsim1\Mpc$~\cite{Ade:2015cva}. Similarly, the CMB distortion gives a slightly milder but nontrivial upper bound on $B_0$ on smaller scales~\cite{Jedamzik:1999bm}.
Combining these constraints, one finds the magnetic field should satisfy  $10^{-14} \G <B_0<10^{-8} \G$ and $1 {\rm pc} < \lambda_0 < 1 \Mpc$ in the case (i) or (ii) (see the blue line in Fig.~\ref{fig5}). These constraints are summarized in Fig.~\ref{fig5}.

\begin{figure}[htbp]
\centering{
\includegraphics[width = 0.45\textwidth]{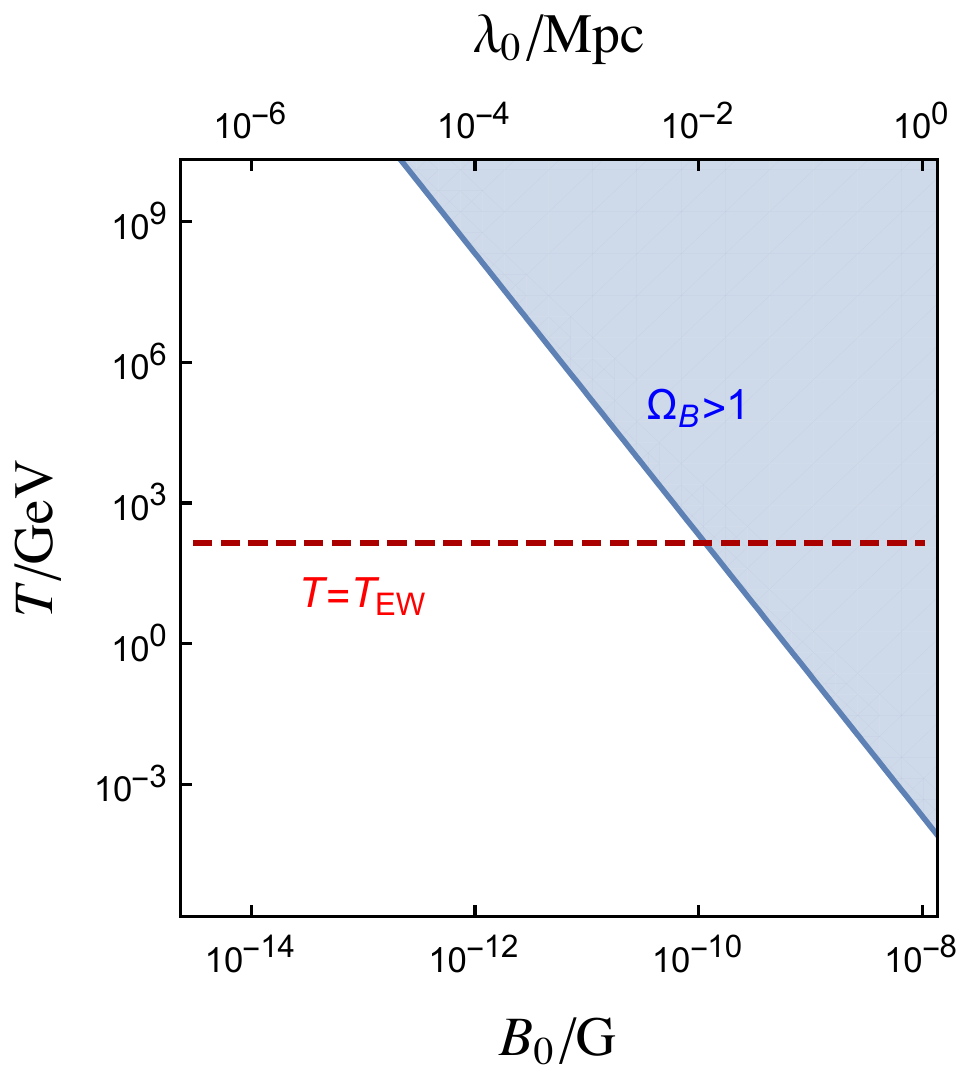}}
\caption{The parameter region where the magnetic fields would dominate the energy density of the Universe $\Omega_B>1$ in the solely inverse cascade case (i) is shown as the shaded region. 
The  dashed line represents the electroweak scale. In the case of $B_0>10^{-10} \G$ or $\lambda_0>10^{-2} \Mpc$, the magnetic fields must be generated or make the transition from the adiabatic evolution to the inverse cascade regime at the lower temperatures than the electroweak phase transition.   }
\label{fig0}
\end{figure}
Finally, the energy fraction of the magnetic field is an increasing function of $T$, 
\begin{equation}
\Omega_B^{\rm IC} (T) = \frac{(B_p^{\rm IC})^2}{2\rho_{\rm tot}} \simeq   6 \times 10^{-9} \left(\frac{T}{10^2 \GeV}\right)^{2/3}
\left(\frac{B_0}{10^{-14}\G}\right)^{4/3}
\left(\frac{\lambda_0}{1\pc}\right)^{2/3}\mathcal{G}_\Omega(T),
\end{equation}
for the case (i) and for $T<T_{\rm TS}$ in the case (ii).
On the other hand, it does not depend on $T$,
$\Omega_B^{\rm AD}(T)=\Omega_B^{\rm IC}(T_{\rm TS})$,
for $T>T_{\rm TS}$ in the case (ii). Here we define
$\mathcal{G}_\Omega(T)\equiv (g_{*}^{\rm tot}(T)/106.75)^{-2/3}$ $(g_{*}^{\rm ch}(T)/80)^{1/3}$ $(g_{*s}(T)/106.75)^{2/3}$. Therefore, the magnetic energy density would overwhelm that of radiation at
\begin{equation}
T>T_{\rm dom} \equiv 2\times 10^2 \GeV \left(\frac{B_0}{10^{-10} \G}\right)^{-2} \left(\frac{\lambda_0}{10^{-2} \Mpc}\right)^{-1} \label{tdom}
\end{equation}
for the case (i) and for $T_{\rm TS}>T_{\rm dom}$ in the case (ii),  
as shown in Fig.~\ref{fig0}. 
In this paper, we do not consider the case where such a magnetic dominated Universe emerged prior to the standard radiation dominated Universe. 
Thus the magnetic fields must be generated or experience the transition from the adiabatic evolution to the inverse cascade 
regime at $T<T_{\rm dom}$. In particular, in the case of $B_0>10^{-10} \G$ or $\lambda_0>10^{-2} \Mpc$, 
we do not expect that there are magnetic fields that undergo the inverse cascade process at the electroweak phase transition.

\section{Chiral anomaly in the Standard Model and baryogenesis from helical magnetic field}
\label{sec3}

Now we study the chiral anomaly in the SM and see how the baryon asymmetry 
is generated through the background helical magnetic field. 
The SM based on the $SU(3)_{C} \times SU(2)_{L} \times U(1)_{Y}$ 
contains three types of gauge bosons [$Y \ {\rm for} \ U(1)_Y, W \ {\rm for} \ SU(2)_L, G \  
 {\rm for} \ SU(3)_C$], three generations of quarks and leptons, 
and the Higgs scalar ($\varphi$). 
The SM fermion currents are known to be anomalous 
due to the coupling to gauge bosons and are not conserved 
even in the massless and free limit \cite{'tHooft:1976up}
\begin{equation}
\nabla_\mu j^\mu_f = C_{\rm y}^f \frac{\alpha_{\rm y}}{4 \pi} Y_{\mu\nu} {\tilde Y}^{\mu\nu}+C_{\rm w}^f \frac{\alpha_{\rm w}}{8\pi} W^a_{\mu\nu}{\tilde W}^{a \mu\nu}+C_{\rm s}^f\frac{\alpha_{\rm s}}{8\pi}G^b_{\mu\nu}{\tilde G}^{b\mu\nu}, \label{anom}
\end{equation}
where $\nabla_\mu$ is the covariant derivative, the current of the Weyl fermion species $\chi_f$ is defined as 
$j^\mu_f \equiv \chi_f^\dagger {\bar \sigma}^\mu \chi_f$ for the left-handed fermions and 
$j^\mu_f \equiv \chi_f^\dagger {\sigma}^\mu \chi_f$ for the right-handed fermions,  
$\alpha_{Y,(W,C)} \equiv g_{1,(2,3)}^2/4 \pi$ are the fine structure constants of 
each gauge symmetry, 
\\
${\tilde X}^{\mu\nu} \equiv (1/2) \epsilon^{\mu\nu\rho\sigma} X_{\rho \sigma}/\sqrt{-{\rm det} (g_{\mu\nu})}= (1/2) a^{-3}(t) \epsilon^{\mu\nu\rho\sigma} X_{\rho \sigma} (X=Y,W,G)$ 
are the dual gauge field strength tensors, and 
the coefficients $C_i^f$ are summarized in Table \ref{tab2}. 
$N_{\rm c}=3, N_{\rm w}=2$ are the number of degrees of freedom of the color and weak 
isospin states of leptons and quarks, respectively, and 
$q_Q=1/6, q_u=2/3, q_d=-1/3, q_L=-1/2$, and $q_e=-1$ are the 
hypercharge of each quarks and leptons.  

\begin{table}[tbp]
\begin{center}
\caption{The coefficients in Eq.~\eqref{anom}. $i=1,2,3$ are the generation indices. }
\begin{tabular}{l||ccc}
&$C_{\rm y}$&$C_{\rm w}$&$C_{\rm s}$ \\ \hline 
$Q^i$ & $N_{\rm c} N_{\rm w} q_Q^2$& $N_{\rm c}$ & $N_{\rm w}$ \\
$u_R^i$ & $-N_{\rm c} q_u^2$ & $0$ & $-1$ \\
$d_R^i$ & $-N_{\rm c} q_d^2$ & $0$ & $-1$ \\
$L^i$ & $N_{\rm w} q_L^2$ & $1$ & $0$ \\
$e_R^i$ & $-q_e^2$ & $0$ & $0$ 
\end{tabular}
\label{tab2}
\end{center}
\end{table}

The second and third terms in the right-hand side of Eq.~\eqref{anom}
induce so-called weak and strong sphaleron processes \cite{Klinkhamer:1984di,McLerran:1990de}. 
The $SU(2)$ and $SU(3)$ gauge theories have degenerate vacua, 
whose Chern-Simons (CS) number ($N_{\rm CS}$) are integers. 
The sphaleron processes are the transition process from one vacuum to another, 
which results in the change of number of quarks and leptons. 
In particular, the weak sphaleron is accompanied by the violation of 
$B$ and $L$ numbers (but not $B-L$), and hence it plays the crucial 
role in many baryogenesis mechanisms \cite{Kuzmin:1985mm,Fukugita:1986hr}.    
At high temperatures above the electroweak scale, 
the sphaleron rate or the CS diffusion rate is estimated by the numerical simulations
as $\Gamma_{\rm w} \simeq 25 \alpha_{\rm w}^5 T$ for the weak sphaleron \cite{Moore:1997sn,Bodeker:1999gx}
and $\Gamma_{\rm s} \simeq 100 \alpha_{\rm s}^5 T$ for the strong sphaleron \cite{Moore:1997im}. 

On the other hand, the vacuum structure of the $U(1)$ gauge theory is trivial, 
and hence the sphaleron-like effect does not occur in the hypercharge sector
in the vacuum.\footnote{However, the thermal fluctuations of the hypermagnetic helicity 
can affect the evolution of the baryon and lepton asymmetry \cite{Long:2013tha,Long:2016uez}. Although the effect is expected to be small, these works leaving it open ended whether this effect makes much difference to the baryogenesis calculations or not. } 
However, as we have seen in the previous section, 
there can be time-varying nontrivial hypermagnetic field with a net helicity in the early Universe, 
which contributes to the anomalous process \cite{Anber:2015yca}.
(See also Refs.~\cite{Joyce:1997uy,Brustein:1998du,Giovannini:1997eg, Giovannini:1999wv,Giovannini:1999by,Bamba:2006km,Bamba:2007hf,Boyarsky:2011uy,Dvornikov:2011ey,Boyarsky:2012ex,Semikoz:2012ka,Tashiro:2012mf,Dvornikov:2012rk,Long:2013tha,Dvornikov:2013bca,Semikoz:2013xkc,Sabancilar:2013raa,Bhatt:2015ewa,Semikoz:2015wsa,Boyarsky:2015faa,Zadeh:2015oqf,Long:2016uez}.) Note that the volume average of  $Y_{\mu\nu} {\tilde Y}^{\mu\nu}$ is proportional to the rate of change of the helicity density ${\dot h}$,
\begin{equation}
\lim_{V\to \infty} \frac{1}{V}\int_V\dd^3 x  
\, Y_{\mu\nu}\tilde{Y}^{\mu\nu}
=2 a^{-3}\, \dot{h}.
\end{equation}
Thus the fermionic currents are not conserved when ${\dot h} \not = 0$ 
and their divergence has a source term proportional to ${\dot h}$.
As a result, at temperatures above the electroweak scale, $B$ (and $L$) is not conserved (while  $B-L$ {\it is} conserved), and the anomalous process can generate the baryon asymmetry of the Universe. 

Now we study the evolution of the asymmetry of the fermions. 
Since the 0th component of the fermionic current represents the net number density 
of the fermion (number density of particle minus antiparticle; $n_f-n_{\bar f}$), 
the evolution equations for the fermion number density-to-entropy ratio
$\eta_f \equiv j_f^0/s$ with $s \equiv 2 \pi^2 g_{*s}T^3/45$ averaged over the present cosmological scales 
as well as that of the Higgs $\varphi$ in the radiation dominated era 
are given by~\cite{Anber:2015yca}
\footnote{
Here we assume that there are no other backreaction effects to the helicity of magnetic fields than the chiral 
magnetic effects (which we have confirmed is does not change our result significantly). It may be true if the helical magnetic fields act as catalyzers in the chiral anomaly. 
Even if there are other unknown back reaction effects, we expect that it is negligible since the asymmetric part of the energy density of baryons are smaller than that of magnetic fields.}, %
\begin{align}
\frac{\partial \eta_{Q^i}}{\partial x}=&-N_{\rm c} N_{\rm w} q_Q^2 \gamma_{\rm y} -N_{\rm c} \gamma_{\rm w} \sum_j (\eta_{Q^j}+\eta_{L^j})-N_{\rm w} \gamma_{\rm s} \sum_j (\eta_{Q^j}-\eta_{u^j}-\eta_{d^j}) \notag \\
&-\sum_j \gamma_{u^{ij}} \left(\frac{\eta_{Q^i}}{6}+\frac{\eta_\varphi}{2}-\frac{\eta_{u^j}}{3}\right)-\sum_j \gamma_{d^{ij}} \left(\frac{\eta_{Q^i}}{6}-\frac{\eta_\varphi}{2}-\frac{\eta_{d^j}}{3}\right), \label{bol1}\\
\frac{\partial \eta_{L^i}}{\partial x}=&-N_{\rm w} q_L^2 \gamma_{\rm y}- \gamma_{\rm w} \sum_j (\eta_{Q^j}+\eta_{L^j})-\sum_j \gamma_{e^{ij}} \left(\frac{\eta_{L^i}}{2}-\frac{\eta_\varphi}{2}-\eta_{e^j}\right), \label{bol2}\\
\frac{\partial \eta_{u^i}}{\partial x}=&N_{\rm c} q_u^2 \gamma_{\rm y}+\gamma_{\rm s} \sum_j (\eta_{Q^j}-\eta_{u^j}-\eta_{d^j})+\sum_j \gamma_{u^{ji}} \left(\frac{\eta_{Q^j}}{6}+\frac{\eta_\varphi}{2}-\frac{\eta_{u^i}}{3}\right), \label{bol3}\\
\frac{\partial \eta_{d^i}}{\partial x}=&N_{\rm c} q_d^2 \gamma_{\rm y}+\gamma_{\rm s} \sum_j (\eta_{Q^j}-\eta_{u^j}-\eta_{d^j})+\sum_j \gamma_{d^{ji}} \left(\frac{\eta_{Q^j}}{6}-\frac{\eta_\varphi}{2}-\frac{\eta_{d^i}}{3}\right), \label{bol4}\\
\frac{\partial \eta_{e^i}}{\partial x}=&q_e^2 \gamma_{\rm y}+\sum_j \gamma_{e^{ji}} \left(\frac{\eta_{L^j}}{2}-\frac{\eta_\varphi}{2}-\eta_{e^i}\right), \label{bol5}\\
\frac{\partial \eta_{\varphi}}{\partial x}=&-\sum_{i,j}\gamma_{u^{ij}} \left(\frac{\eta_{Q^i}}{6}+\frac{\eta_\varphi}{2}-\frac{\eta_{u^j}}{3}\right)+\sum_{i,j}\gamma_{d^{ij}} \left(\frac{\eta_{Q^i}}{6}-\frac{\eta_\varphi}{2}-\frac{\eta_{d^j}}{3}\right)+\sum_{i,j}\gamma_{e^{ij}} \left(\frac{\eta_{L^i}}{2}-\frac{\eta_\varphi}{2}-\eta_{e^j}\right). \label{bol6}
\end{align}
Here we take into account the Yukawa interactions, and 
the time variable 
$x$ is defined as
$x\equiv \sqrt{90/\pi^2g_* } M_{\rm pl}/T$. 
Note that in the radiation dominated era, $H=1/2t$ and $3 H^2 M_{\rm pl}^2 = 
(\pi^2 g_*/30) T^4$.  
The dimensionless interaction rates $\gamma$ are given by 
$\gamma_{\rm w}=\Gamma_{\rm w}/T \simeq 25 \alpha_{\rm w}^5, \gamma_{\rm s}=
\Gamma_{\rm s}/T\simeq 100 \alpha_{\rm s}^5$, 
and $\gamma_{u (d,e)^{ij}}=\Gamma_{u (d,e)^{ij}}/T = |y_{u (d,e)}^{ij}|^2/8\pi$
with $y_{u (d,e)}^{ij}$ being the Yukawa coupling matrices for 
up-type quarks, down-type quarks, and electron-type leptons, respectively. 
Note that $i=1,2,3$ runs the generation. 
$\gamma_{\rm y}$ in the source term from the helical magnetic field is defined as 
\begin{equation}
\gamma_{\rm y}\equiv a^{-3} \frac{\alpha_{\rm y}}{2 \pi s}\frac{\dot h}{T}. 
\end{equation}
(See Appendix \ref{ap2} for the numerical values of these constants.)
By solving the 16 evolution equations Eqs.~\eqref{bol1}-\eqref{bol6} 
from the emergence of thermal plasma of SM particles or magnetogenesis, whichever comes later,\footnote{
If magnetogenesis took place before reheating, modified evolution equations before reheating should be used. However, the resultant baryon asymmetry is determined by the dynamics around the electroweak scale, as we see below. Thus we do not explore it assuming that reheating took place before the electroweak phase transition. }
to the electroweak scale $T=T_{\rm f}\simeq 140$ GeV
at which  the weak sphaleron process shuts off \footnote{
At temperature below the electroweak scale, 
the anomalous process from the helical magnetic field does not violate 
$B$ (and $L$) and hence $B$ (and $L$) asymmetries do not change anymore. 
}~\cite{Burnier:2005hp}, 
we can estimate the baryon asymmetry 
\begin{equation}
\eta_B \equiv \frac{1}{3} \sum_i (\eta_{Q^i}+\eta_{u^i}+\eta_{d^i})
\end{equation}
generated by this process.  
Note that here we take into account the Yukawa interaction and hence chemical 
potential for the Higgs field, as discussed in the introduction. It relates all the chemical potential of quarks and leptons 
nontrivially. Therefore we must solve all 16 evolution equations simultaneously to 
acquire the precise results.
In the case where the present magnetic field is maximally helical (see Appendix \ref{ap1} for its definition), the source term from the helical magnetic field is given by 
\begin{align}
\gamma_{\rm y} &\simeq  1.7 \times 10^{-26} \mathcal{C} \left(\frac{B_0}{10^{-14} {\rm G}}\right)^2 \left(\frac{\lambda_0}{10^{-6} {\rm Mpc}}\right)^{-1}
\times \left\{\begin{array}{ll} 
\left(\dfrac{T}{1{\rm GeV}}\right)^{4/3} & \quad \text{for} \quad T<T_{\rm TS} \\
\left(\dfrac{T_{\rm TS}}{1{\rm GeV}}\right)^{4/3} &\quad  \text{for} \quad T>T_{\rm TS}
\end{array}\right. \notag \\
& \simeq   1.1 \times 10^{-2}\mathcal{C} \left(\frac{B_0}{10^{-14} {\rm G}}\right)
\times \left\{\begin{array}{ll} 
x^{-4/3} & \quad \text{for} \quad T<T_{\rm TS} \\
x_{\rm TS}^{-4/3} & \quad \text{for} \quad T>T_{\rm TS}
\end{array}\right.  \label{ICgy}
\end{align}
where $x_{\rm TS}\equiv \sqrt{90/\pi^2 g_*} M_{\rm pl}/T_{\rm TS}$.  
Here we used Eq.~\eqref{presentMF} and  introduce a numerical factor $\mathcal{C}$ to take into account the uncertainty caused by the approximated equations which have been used so far 
such as $\lambda_B \simeq v_A t$ or $\sigma \simeq 100T$.
We expect that the uncertainty is at most $0.1 \lesssim {\mathcal{C}} \lesssim 10$. 
In this case, the baryon asymmetry which is consistent with the present observation can be generated as we will see in detail in the next section. 

One may wonder if the weak sphaleron washes out the 
baryon asymmetry generated by this process and the resultant asymmetry 
is exponentially suppressed, since this mechanism does not generate $B-L$ asymmetry.\footnote{ Indeed, one can explicitly show $\partial_t \eta_{\rm B-L}=0$ from Eqs.~\eqref{bol1}-\eqref{bol6}.}
However, it is not the case for two reasons. 
First, the washout mechanism significantly works only after all the Yukawa interactions as well as the weak sphaleron process become active. 
But electron Yukawa coupling $y_{e^{11}}\sim 10^{-6}$ is so small that 
it becomes effective only at temperatures below $T=10^{5-6}$ GeV, or $x>10^{12-13}$. Therefore, relatively large baryon asymmetry can be produced at a temperatures above $T=10^{5-6}$ GeV. 
Second, when the electron Yukawa interaction becomes effective, 
the baryon asymmetry would decay exponentially if there is no source term from the helical magnetic field. However, with the aid of the source term, the decay of the baryon asymmetry significantly slows down and it is no longer the exponential damping but is only at most a power law of $T$.
Although the precise estimate can be done only by solving all the 16 evolution equations numerically, the qualitative behavior of the evolution of the baryon asymmetry can be understood 
by examining the following simplified equation, 
\begin{equation}
\frac{\partial \eta_B(x)}{\partial x} = \gamma_y(x) - \gamma_{e^{11}} \eta_B(x).  
\end{equation}
Note that the electron Yukawa interaction is the last piece required to activate the wash-out  effect of the baryon asymmetry
as explained above. 
At later times $x \gg \gamma_{e^{11}}^{-1} \approx 3\times 10^{12}$, the source term and damping force equilibrate, and hence the baryon asymmetry has an attractor solution,
\begin{equation}
\eta_B(x) \sim \frac{\gamma_y(x)}{\gamma_{e^{11}}}. \label{anal}
\end{equation}
In other words, although the sphaleron and Yukawa interactions try to damp the baryon asymmetry 
exponentially, the source term from the helical magnetic field prevents it by continuously producing $\eta_B$.  
Equation~\eqref{anal} predicts the following simple behavior of $\eta_B$
which will be confirmed by numerical calculations in the next section:
If the inverse cascade process takes place above the electroweak scale, namely the case (i) or the case (ii) with $T_{\rm f} \simeq T_{\rm EW} <T_{\rm TS}$, the baryon asymmetry evolves as $\eta_B\, \propto \gamma_y \propto  x^{-4/3}$ for $\gamma_{e^{11}}^{-1}<x<x_{\rm f}$. If the helical magnetic field adiabatically evolves, namely the case (ii) with $T_{\rm f} \simeq T_{\rm EW} >T_{\rm TS}$ or the case (iii), $\eta_B$ becomes constant.

\section{Numerical result} 
\label{sec4}

Here we examine the scenario numerically and give quantitative evaluations. 
We assume that the magnetic field indicated by the blazar observation 
is maximally helical~\footnote{It should be noted if magnetic fields are partially helical at their generation, the helical part decays slower than the nonhelical part due to the inverse cascade process and they eventually reach the maximal helical state~\cite{Banerjee:2004df, Saveliev:2013uva}. Therefore our assumption that the present magnetic fields are maximally helical is valid for a broad class of initial conditions. Furthermore, since only the helical part of the magnetic field contribute to produce the baryon asymmetry through the chiral anomaly, we do not need to regain the nonhelical part which decays during the evolution.} and generated before the electroweak phase transition.\footnote{We also assume that reheating took place before the electroweak phase transition.} 
The initial temperature in our calculation $T_{\rm ini}$ is understood as the temperature at which magnetogenesis finishes or the reheating temperature in the case of inflationary magnetogenesis.  
In the following we neglect the running of gauge and Yukawa couplings 
since they run only logarithmically with respect to the energy scales. 

\begin{figure}[htbp]
\centering{
\includegraphics[width = 0.65\textwidth]{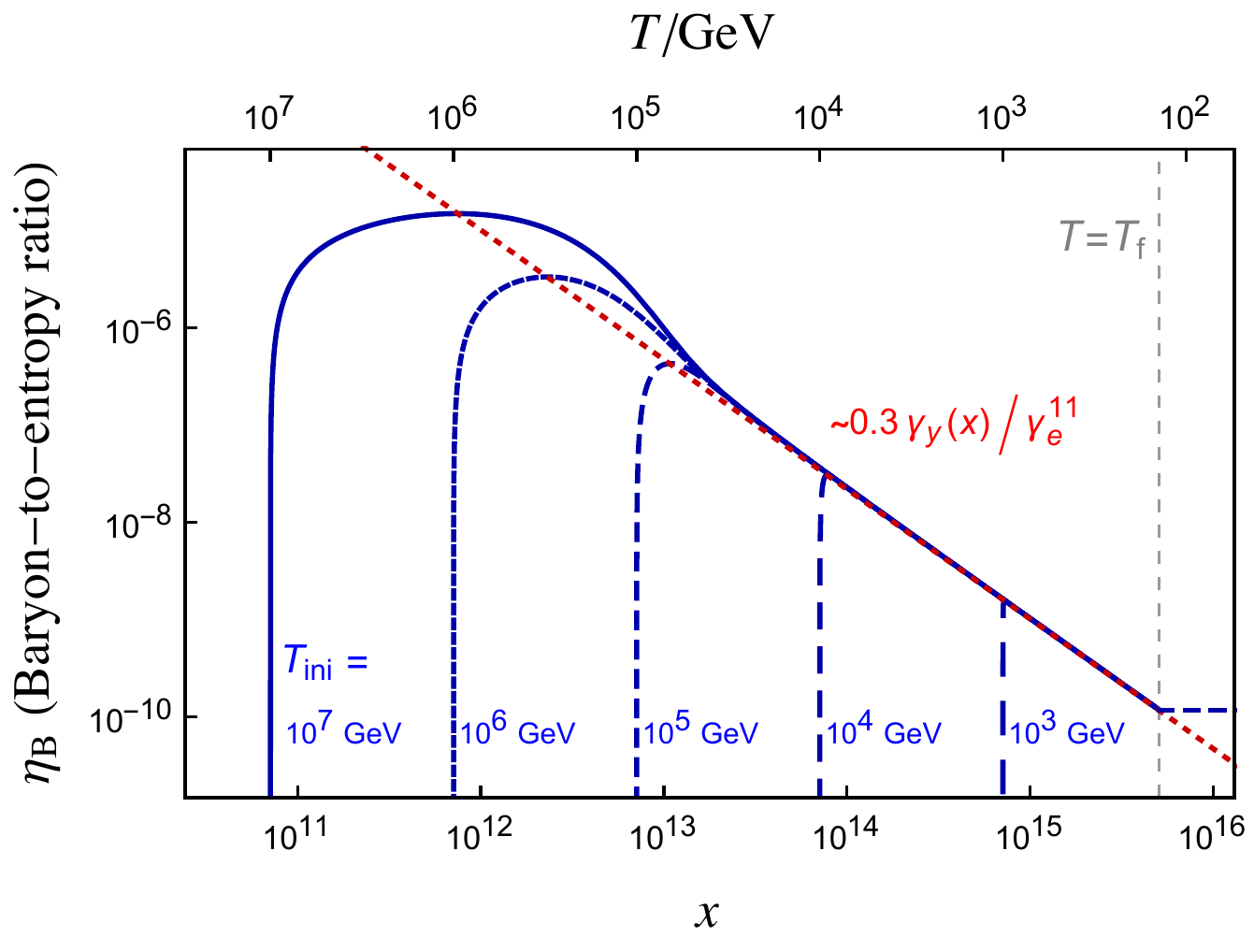}}
\caption{The evolution of baryons asymmetry for 
$B_0=10^{-13} \G$ in the solely inverse cascade case ($T_{\rm ini}<T_{\rm TS}$) is shown. The horizontal axis denotes $x\equiv \sqrt{90/\pi^2g_* } M_{\rm pl}/T$. The initial temperature 
is taken as $T_{\rm ini}=10^7 \GeV, \   10^6 \GeV, \ 10^5 \GeV,  \  10^4  \ {\rm GeV}$, and $10^3$ GeV from left to right. The  dotted line shows that the asymptotic behavior at $T<10^5$ GeV is well fitted by 
$0.3 \gamma_{\rm y}/\gamma_{e^{11}}$. } 
\label{fig1}
\end{figure}
\begin{figure}[htbp]
\centering{
\includegraphics[width = 0.65\textwidth]{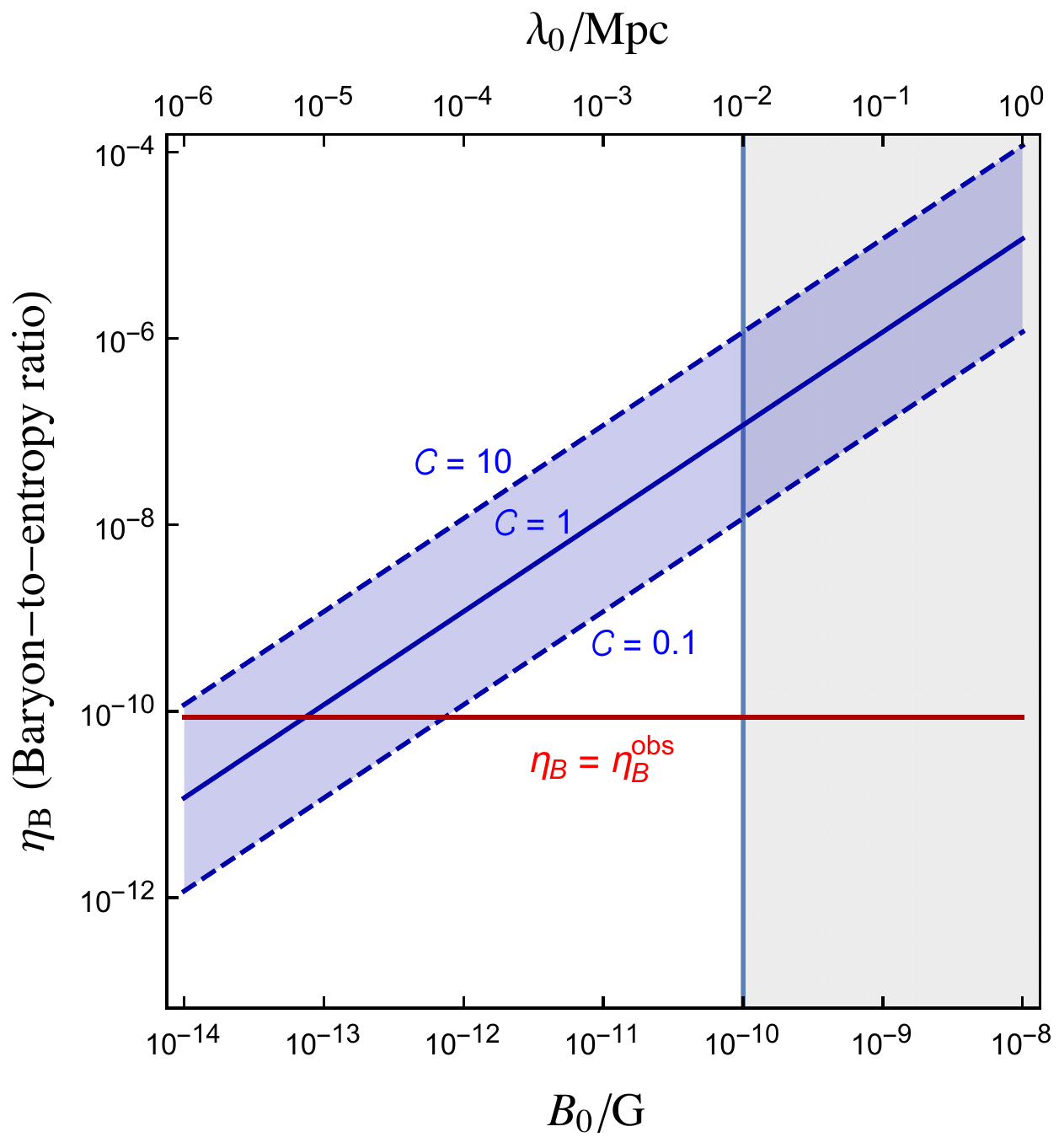}}
\caption{The final baryon asymmetry with respect to the present magnetic field is shown in the solely inverse cascade case (i) and the transition case with $T_{\rm TS}<T_f$.
The blue shaded region represents the theoretical uncertainty, 
which is parametrized by the parameter $\mathcal{C}$.
The  line shows the observed baryon asymmetry. The gray shaded region is disfavored from the condition $\Omega_B>1$ at $T>T_{\rm EW}$ (see Fig.~\ref{fig0}).  }
\label{fig2}
\end{figure}
Figure~\ref{fig1} shows the time evolution of the baryon asymmetry 
for $B_0=10^{-13}$ G (and $\lambda_0 = 10$ pc) in the case (i) (the solely inverse cascade case; $T_{\rm TS}>T_{\rm ini}$).  
$\gamma_{\rm y}(x)$ is evaluated by Eq.~\eqref{ICgy} with  
${\mathcal C}=1$. We take the initial temperature as $T_{\rm ini}=10^7 \  {\rm GeV}, 10^6 \ {\rm GeV}, 10^5  \ {\rm GeV}, 10^4  \ {\rm GeV}$, 
and $10^3$ GeV with the initial condition $\eta_f=0$. We can see 
that $\eta_B$ evolves as
\begin{equation}
\eta_B(x) \simeq 0.3 \frac{\gamma_y(x)}{\gamma_{e^{11}}} \simeq 10^{-10} \mathcal{C} \left(\frac{B_0}{10^{-13} \G}\right)\left(\frac{x}{x_{\rm f}}\right)^{-4/3}, \end{equation}
where $x_{\rm f} \equiv \sqrt{90/\pi^2 g_*} M_{\rm pl}/T_{\rm f}$.
Therefore the numerical results show an excellent agreement with the analytic estimate in the previous section [Eq.~\eqref{anal}].
We can also see that the resultant asymmetry is  independent of the initial time, if the initial temperature is sufficiently larger than the electroweak scale.
This is because the source term and damping force from electron Yukawa interaction determine the final asymmetry as the attractor solution Eq.~\eqref{anal}. Note that the case (ii) with $T_{\rm ini}>T_{\rm TS}>T_{\rm f} \simeq T_{\rm EW}$ shows the same behavior.

Figure~\ref{fig2} illustrates the resultant baryon asymmetry 
for varying $B_0$ in the case where the helical magnetic field enters the inverse cascade regime before the electroweak phase transition ($T_{\rm TS}>T_{\rm f}\simeq T_{\rm EW}$). 
We take into account the theoretical uncertainty of the source term [Eq.~\eqref{ICgy}]
by means of the parameter $0.1<\mathcal{C}<10$.
We can see that for $10^{-14} \G <B_0<10^{-12} \G$, the present baryon asymmetry 
can be explained within the theoretical uncertainties. 
In contrast, the region $10^{-12} \G <B_0<10^{-10}\G$ predicts the over production of baryon asymmetry. Such parameter regions are disfavored for the case (i) or the case (ii) with $T_{\rm TS}>T_{\rm EW}$. Consequently, if future observations suggests $B_0>10^{-12} \G$ satisfying Eq.~\eqref{presentMF}, the magnetogenesis or the transition from the adiabatic evolution to the inverse cascade regime must take place after the electroweak phase transition. 

\begin{figure}[htbp]
\centering{
\includegraphics[width = 0.65\textwidth]{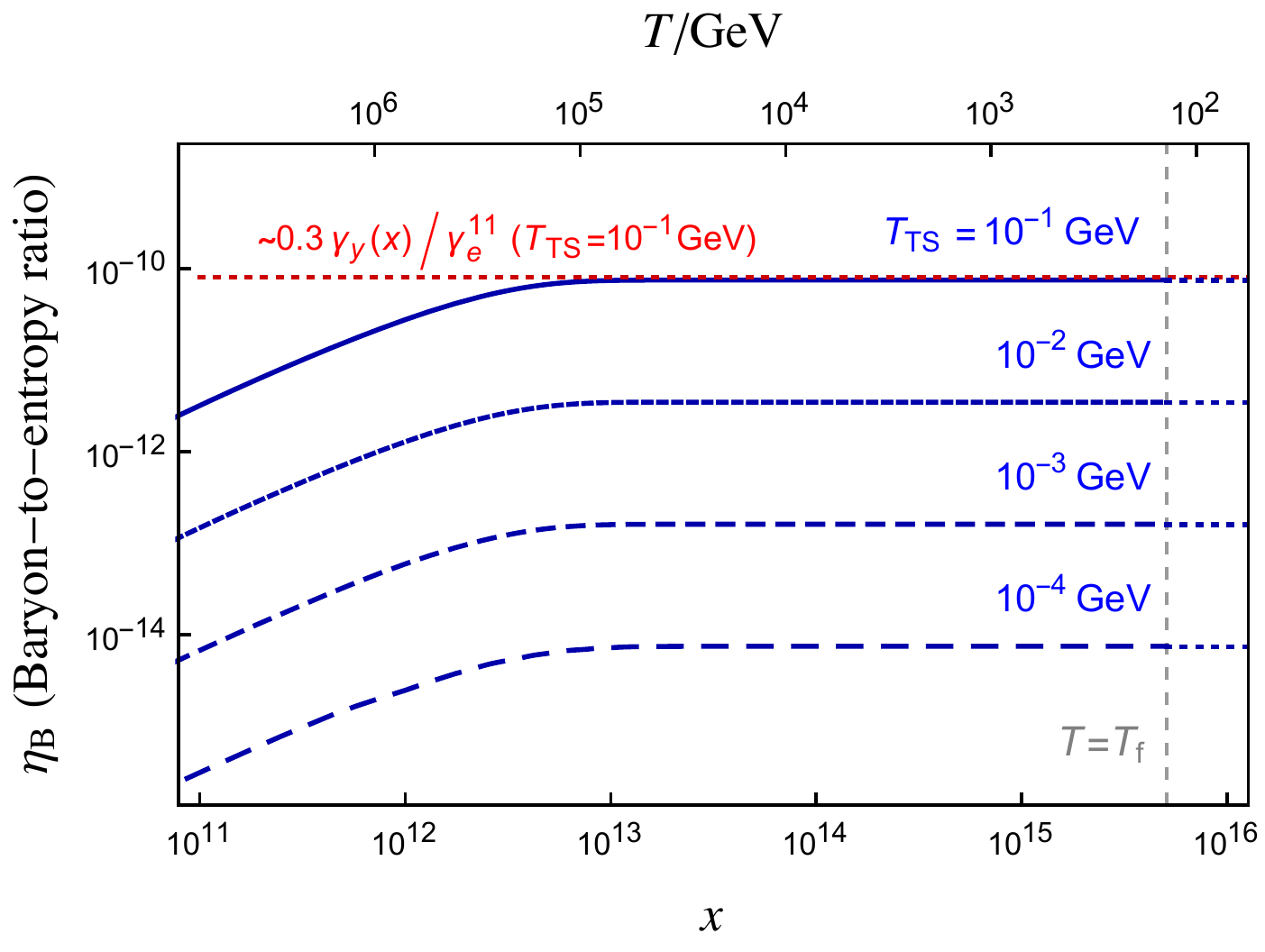}}
\caption{The evolutions of baryon asymmetry for $B_0=10^{-9} \G$ with various 
$T_{\rm TS}(<T_{\rm EW})$ are shown. The transition temperature is taken as 
$T_{\rm TS}=10 \GeV, 1 \GeV, 10^{-1} \GeV, 10^{-2} \GeV$, and $10^{-3} \GeV$.  dashed line 
shows that the asymptotic behavior at $T<10^5 \GeV$ for $T_{\rm TS}=10^{-2} \GeV$ 
is well fitted by 
$0.3 \gamma_{\rm y}/\gamma_{e^{11}}$.}
\label{fig3}
\end{figure}
The baryon overproduction problem for large $B_0$ can be relaxed by
supposing the transition case (ii) with $T_{\rm TS}<T_{\rm f} \simeq T_{\rm EW}$. 
Figure~\ref{fig3} shows the time evolution of the baryon asymmetry with $B_0=10^{-9} \G$ with various $T_{\rm TS}$ in that case. 
We take the transition temperature as 
$T_{\rm TS}= 10^{-1} \GeV, 10^{-2} \GeV, 10^{-3} \GeV$, and $10^{-4} \GeV$. 
We can see that the baryon asymmetry saturates at $T \simeq 10^5 \GeV$, 
as suggested by the attractor behavior [Eq.~\eqref{anal}]. 
As is the case of $T_{\rm TS}>T_{\rm f} \simeq T_{\rm EW}$, 
the asymptotic behavior of the baryon asymmetry can be expressed as 
\begin{equation}
\eta_B(x) \simeq 0.3 \frac{\gamma_y(x)}{\gamma_{e^{11}}} \simeq 10^{-10} \mathcal{C} \left(\frac{B_0}{10^{-13} \G}\right)\left(\frac{x_{\rm TS}}{x_{\rm f}}\right)^{-4/3}. 
\end{equation} 
In particular, even if the magnetogenesis or reheating took place 
before the electroweak phase transition and the present magnetic field is stronger than $10^{-12}\G$, 
$\eta_B=\eta_B^{\rm obs}\simeq 0.86 \times 10^{-10}$ is realized for
\begin{equation}
T_{\rm TS}\simeq 10^{-1} \GeV \mathcal{C}^{-3/4} \left(\frac{B_0}{10^{-9} \G}\right)^{-3/4}\equiv T_{\rm TS}^{\rm b}.
\label{Ttrb}
\end{equation}
If the transition is too early $T_{\rm TS}>T^{\rm b}_{\rm TS}$ or too late $T_{\rm TS}<T^{\rm b}_{\rm TS}$, however, the resultant baryon asymmetry becomes larger or smaller than the observed value, respectively. 
Note that we require that magnetic fields never dominates the 
energy density of the Universe, $T_{\rm TS}<T_{\rm dom}$ [Eq.~\eqref{tdom}]. 
In order to be consistent, $T_{\rm TS}^{\rm b}<T_{\rm dom}$ must be satisfied, 
which turns to be the upper bound on the strength of the present magnetic field
that can explain the baryon asymmetry of our Universe, 
\begin{equation}
B_0<10^{-9} \G \times \mathcal{C}^{1/3}. 
\end{equation}
Here Eq.~\eqref{presentMF} is used.

For completeness, let us make a comment on the solely adiabatic case (iii).
Roughly speaking, this case can be seen as the special case of the transition case (ii) whose transition temperature $T_{\rm TS}$ is lower than the present temperature $T_{\rm 0}\approx 2\times10^{-13}\GeV$.
As seen in Fig.~\ref{fig3}, in the limit $T_{\rm TS}\to 0$, the baryon asymmetry becomes negligible. The only difference between the case (iii) and the case (ii) with $T_{\rm TS}<T_{\rm 0}$ is that the present correlation length $\lambda_0$ can be longer in the case (iii) than the case (ii), because Eq.~\eqref{presentMF} is not applied. However, since $\dot{h}\propto \lambda_B^{-1}$, the longer correlation length leads to smaller baryon asymmetry. Therefore the solely adiabatic evolution case (iii) cannot explain the observed baryon asymmetry.

\section{Summary and Discussion}

In this paper, we study the generation of baryon asymmetry from the helical magnetic field in the primordial Universe through the chiral anomaly. In this mechanism, the time-varying helicity of hypermagnetic field spontaneously
breaks $T$ symmetry as well as $C$ and $CP$ symmetry. The chiral anomaly then breaks $B$ symmetry, and the source term from the time-varying helicity prevents the system from entering the complete thermal equilibrium. 
As a result, the Sakharov's condition \cite{Sakharov:1967dj} is satisfied within the SM 
and the baryon asymmetry can be generated.

We assume that there exist helical magnetic fields with negative sign whose 
typical strength and correlation length are $10^{-14} \G < B_0 < 10^{-8} \G$ and $1 {\rm pc} < \lambda_0 < 1 \Mpc$ in the present Universe, satisfying Eq.~\eqref{presentMF},  as the observations indicate and they started to undergo the inverse cascade process before the electroweak phase transition. 
Here we take into account the MHD effect on the magnetic field evolution and all relevant particle interactions including the Yukawa interaction carefully. It is found that the present baryon asymmetry $\eta_B\simeq 10^{-10}$  can be generated 
for $10^{-14} \G < B_0 < 10^{-12} \G$ allowing for the theoretical uncertainties. On the other hand, for stronger helical magnetic fields with $10^{-12} \G < B_0 < 10^{-10} \G$, this mechanism basically causes  the overproduction of baryon asymmetry and hence such strength of present magnetic fields is ruled out. 
The case with $10^{-10} \G < B_0 < 10^{-8} \G$ is already excluded due to the unwanted 
magnetic field domination of the Universe.

\begin{figure}[htbp]
\centering{
\includegraphics[width = 0.65\textwidth]{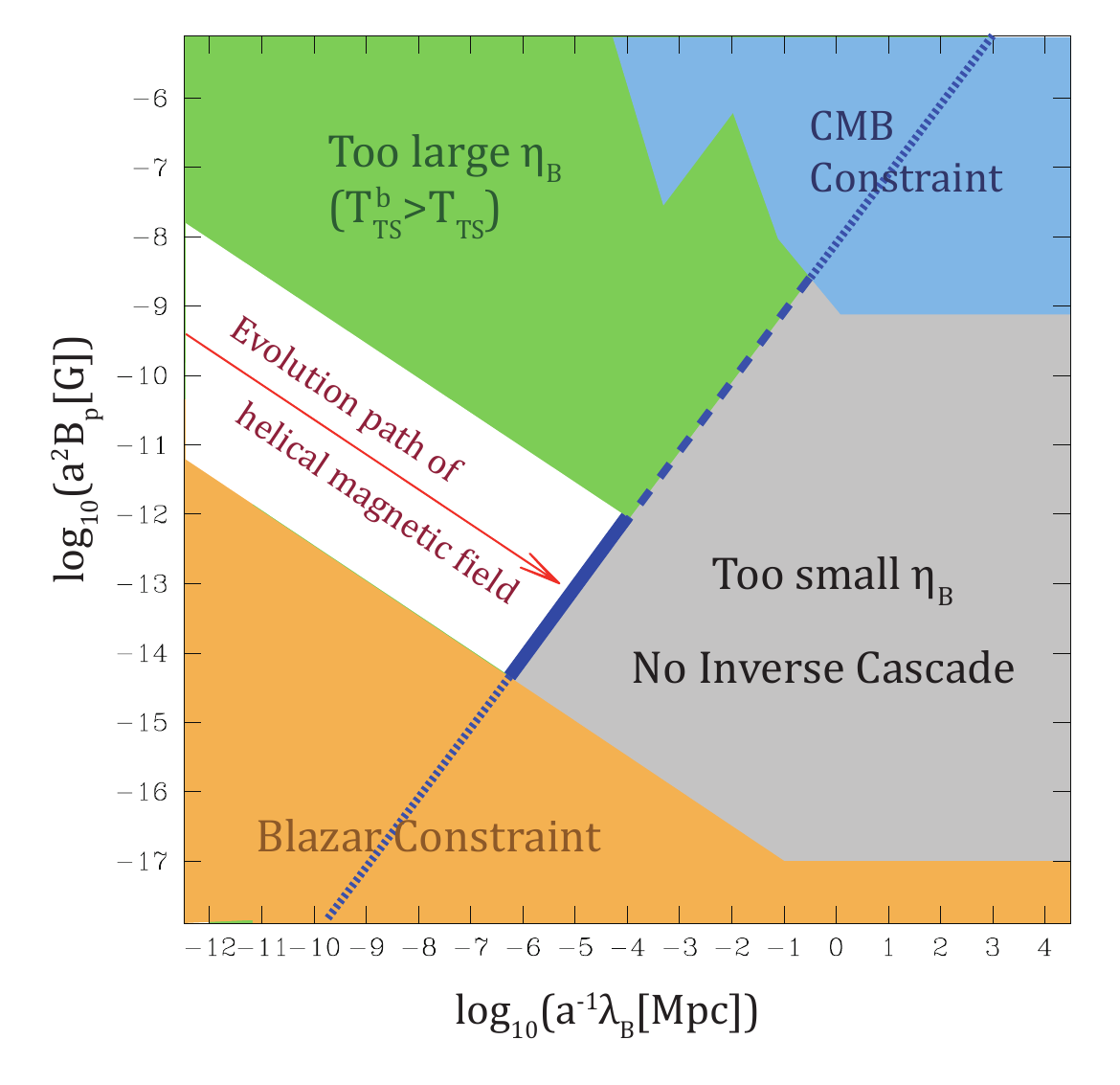}}
\caption{The constraints on the comoving strength and the comoving correlation length of the helical magnetic fields. The blue and orange shaded region show the constraints from the CMB and blazar observations, respectively. In the gray shaded region, although the magnetic fields are allowed to exist, the sufficient baryon asymmetry cannot be generated via the chiral anomaly. In the green region, the helical magnetic field causes the overproduction of baryon asymmetry unless the transition from the adiabatic evolution into the inverse cascade is sufficiently late ($T_{\rm TS}\le T_{\rm TS}^{\rm b}$). The white region shows the window in which the observed baryon asymmetry can be successfully generated. The  arrow represents the evolution path of the helical magnetic field undergoing the inverse cascade process  and eventually reaches the blue thick line [Eq.~\eqref{presentMF}].}
\label{fig5}
\end{figure}

Our result is summarized in Fig.~\ref{fig5}. Interestingly, the favored strength of the helical magnetic field is larger than the observational lower bound by the factor of $\mathcal{O}(1)-\mathcal{O}(10^2)$. Therefore the helical magnetic field which undergo the inverse cascade and generate the baryon asymmetry is expected to be tested in the future observation. Furthermore, as shown in the  line in Fig.~\ref{fig5}, the evolution path of the magnetic field is predicted and hence it would be interesting to target such relationship between the strength and the correlation length for  high-$z$ observations.

The baryon overproduction problem for $B_0>10^{-12} \G$ can be relaxed if the magnetic fields evolved adiabatically 
before the electroweak phase transition and entered the inverse cascade regime at a 
time after that. If the temperature of this transition is $T_{\rm TS}^{\rm b}$ (Eq.~\eqref{Ttrb}), 
the present baryon asymmetry is explained. 
In this case, the present magnetic fields whose strength is up to $B_0<10^{-9} \G$
have a chance to explain the baryon asymmetry of our Universe. 
However, such late-time transitions are generally difficult to be realized for the following reasons. 

In this paper we do not specify the generation mechanism of the helical magnetic field. However, even if we do not assume a specific model, by focusing on a class of magnetogenesis mechanisms and introducing the temperature $T_{\rm gen}$ at which the magnetic field was generated\footnote{Here we focus on the magnetogenesis mechanisms taking place after reheating.}, we can further explore the scenario. 
If the helical magnetic fields are generated 
by a process which occurs within the Hubble horizon, the parameter region is constrained.
By using the temperature $T_{\rm gen}$,
one can rewrite Eq.~\eqref{l case II} as
\begin{align}
T_{\rm TS} \simeq 1{\rm GeV} \left(\frac{\lambda_B^{\rm AD}}{H^{-1}} (T_{\rm gen})\right)^{-3/2}
\left(\frac{T_{\rm gen}}{10^{2}\GeV} \right)^{3/2}
\left(\frac{B_0}{10^{-11}\G} \right)^{3/2}. 
\label{TIC lower bound}
\end{align}
Note that $\lambda_B$ at the generation time $t_i$ cannot exceed the Hubble radius $H^{-1}(T_{\rm gen})$ in the case of such a generation mechanism and that gives the lower bound on $T_{\rm TS}$. Then one obtains the maximum $B_0$ to produce $\eta_B^{\rm obs}$  as
\begin{equation}
B_0 \simeq 1.7\times 10^{-11}\G \, \mathcal{C}^{-\frac{1}{3}} 
\left(\frac{\lambda_B^{\rm AD}}{H^{-1}} (T_{\rm gen})\right)^{\frac{2}{3}}
\left(\frac{T_{\rm gen}}{10^{2}\GeV} \right)^{-\frac{2}{3}}.
\end{equation}
Consequently, as long as a process taking place within the horizon is concerned, 
a magnetogenesis mechanism before the electroweak phase transition 
that results in the present strength of magnetic fields $B_0>10^{-11} \G$ is ruled out 
due to baryon overproduction.
Here, the transition temperature $T_{\rm TS}$ is roughly estimated as
\begin{equation}
T_{\rm TS} \sim 2.3\, \frac{g_{*s}(T_{\rm IC})/g_{*s}(T_{\rm gen})}{\sqrt{g_{*}^{\rm tot}(T_{\rm TS})g_{*}^{\rm ch}(T_{\rm TS})}}\ \frac{B_p(T_{\rm gen}) \Mpl}{\lambda_B(T_{\rm gen}) T_{\rm gen}^3}. \label{TM def}
\end{equation}
We stress again that these restrictions may not be applicable to inflationary magnetogenesis.  

To the best of our knowledge, no magnetogenesis mechanism which consistently produces strong and large-scale helical magnetic fields  has been established~\cite{Vachaspati:2001nb,Copi:2008he,Demozzi:2009fu, Barnaby:2012tk,Fujita:2012rb,Fujita:2013qxa,Fujita:2014sna,Ferreira:2014hma}. For instance, the  natural inflation model or its relatives naturally generate the helical magnetic field by introducing a coupling between the U(1) gauge field and the axion \cite{Caprini:2014mja,Anber:2006xt,Fujita:2015iga}, but the produced magnetic field is too weak to satisfy the observational lower bound in a minimum setup. However, only simplest possibilities have been explored so far, and many other studies are need to be done. 
Our result motivates future additional work on helical magnetogenesis mechanisms. 

For other theoretical aspects of our scenario, several issues also remain. 
One is that there remain theoretical uncertainties in the numerical parameters in the model, parametrized $\mathcal{C}$ in this paper. 
More precise determination of the parameters are required to give more precise 
prediction of the baryon asymmetry. 
This will be accomplished by the further cosmological MHD studies with a concrete initial magnetic spectrum. 
The other is that we assume that the evolution of magnetic fields does not receive 
any effects from the baryon asymmetry generation. 
If there are some effects, they may give further insights on baryogenesis as well as the 
magnetogenesis mechanisms. 
We also assumed that the weak sphaleron and the source term from the helical hypermagnetic 
field switch off instantly and simultaneously at $T=T_{\rm f}$. In order to determine the 
resultant baryon asymmetry precisely, the validity of this assumption should be also 
examined.

\section*{Acknowledgments}
The authors are grateful to M.~Anber, A.~J.~Long, M.~Peskin, E.~Sabancilar, H.~Tashiro, 
T.~Vachaspati and F.~Wilczek for 
helpful discussions and comments. 
The work of T.F.~has been supported in part by the JSPS (Japan Society for the Promotion of Science) Postdoctoral Fellowships for Research Abroad (Grant No. 27-154). 
K.K.~acknowledge support from the DOE for this work under Grant No. DE-SC0013605.

\appendix

\section{The helicity of magnetic fields \label{ap1}}

In this appendix, we describe the helicity of magnetic fields defined in Eq.~\eqref{H def}.
To illustrate its nature, we introduce the following decomposition of the vector potential:
\begin{equation}
 A_i(t, \bm{x})
 = 
 \sum_{\pm} \int \frac{{\rm d}^3 k}{(2\pi)^3} 
 e^{i \bm{k \cdot x}} e_{i}^{(\pm)}(\hat{\bm{k}}) 
 \left[ a_{\bm{k}}^{(\pm)} \mcA_{(\pm)}(k,t) 
  + a_{-\bm{k}}^{(\pm) \dag} \mcA_{(\pm)}^*(k,t) \right]
\,,
\label{quantization}
\end{equation}
where $e^{(\pm)}_i(\hat{\bm{k}})$ are the right/left-handed polarization vectors which satisfy $k_i e_i^{(\pm)}(\hat{\bm{k}})=0$ and $\epsilon_{ijl} k_j e_l^{(\pm)}(\hat{\bm{k}})=\mp i k e_i^{(\pm)}(\hat{\bm{k}})$,
and 
$a_{\bm{k}}^{(\pm) \dag}, a_{\bm{k}}^{(\pm)}$
are the creation/annihilation operators which satisfy the usual commutation relation, $[a^{(\lambda)}_{\bm{k}},a^{(\sigma) \dag}_{-\bm{k}'}]
= (2\pi)^3\delta(\bm{k}+\bm{k}')\delta^{\lambda \sigma}$.
With this decomposition, one can show that the helicity density is written as
\begin{equation}
h=
\int \frac{\dd^3 k}{(2\pi)^3}\, k \Big( |\mcA_+|^2 - |\mcA_-|^2\Big).
\label{def helicity}
\end{equation}
Thus the helicity (density) represents the breaking of the parity symmetry.
Magnetic fields with $h\not = 0$ is called helical magnetic field and 
those with either polarization is negligible compared with the other, namely $|\mcA_-|\gg|\mcA_+|$ or $|\mcA_+|\gg|\mcA_-|$, are said to be maximally helical.

\section{Numerical constants \label{ap2}}

Here we summarize the numerical values of gauge and Yukawa couplings we have used in our numerical calculations; 

\begin{align}
&\alpha_{\rm y}\approx 0.017, \quad \alpha_{\rm w}\approx 0.033, \quad \alpha_{\rm s} \approx 0.11,  \\
& y_{u}^{ij} \approx \left(\begin{array}{ccc} 1.1 \times 10^{-5} & 0 & 0 \\ 0 & 7.1 \times 10^{-3} & 0 \\ 0&0& 0.94\end{array} \right), \\
&y_d^{ij}\approx\left(\begin{array}{ccc} 2.7 \times 10^{-5} & 6.3 \times 10^{-6}  & 2.4 \times 10^{-7}\\ 1.2 \times 10^{-4} & 5.4 \times 10^{-4} & 2.2 \times 10^{-5} \\ 8.3 \times 10^{-5} &9.8 \times 10^{-4}& 2.4 \times 10^{-2} \end{array} \right),\\
&y_{e}^{ij}\approx\left(\begin{array}{ccc} 2.8 \times 10^{-6} & 0 & 0 \\ 0 & 5.8 \times 10^{-4} & 0 \\ 0&0& 1.0 \times 10^{-2} \end{array} \right). 
\end{align}


\end{document}